# Phonon dynamics and thermal conductivity of PtSe$_2$ thin films: Impact of crystallinity and film thickness on heat dissipation

*Alexandros El Sachat\*[1], Peng Xiao[1,2], Davide Donadio[3,4], Frédéric Bonell[5], Marianna Sledzinska[1], Alain Marty[5], Céline Vergnaud[5], Hervé Boukari[5], Matthieu Jamet[5], Guillermo Arregui[1], Zekun Chen[3,4], Francesc Alzina[1], Clivia M. Sotomayor Torres[1,6], Emigdio Chavez-Angel\*[1]*

[1] Catalan Institute of Nanoscience and Nanotechnology (ICN2), CSIC and BIST, Campus UAB, Bellaterra, 08193 Barcelona, Spain

[2] Departamento de Física, Universidad Autónoma de Barcelona, Bellaterra, 08193 Barcelona, Spain

[3] Department of Chemistry, University of California, Davis, California 95616, United States

[4] Ikerbasque, Basque Foundation for Science, E-48011 Bilbao, Spain

[5] Université Grenoble Alpes, CNRS, CEA, Grenoble INP, IRIG-Spintec, 38054 Grenoble

[6] ICREA, Passeig Lluis Companys 23, 08010 Barcelona, Spain

\*Corresponding authors: alexandros.elsachat@icn2.cat, emigdio.chavez@icn2.cat

We present a comparative investigation of the influence of crystallinity and film thickness on the acoustic and thermal properties of 2D layered PtSe$_2$ thin films of varying thickness (0.6 – 24 nm) by combining a set of experimental techniques, namely, frequency domain thermo-reflectance, low-frequency Raman and pump-probe coherent phonon spectroscopy. We find a 35% reduction in the cross-plane thermal conductivity of polycrystalline films with thickness larger than 12 nm compared to the crystalline films of the same thickness due to phonon grain boundary scattering. Density functional theory calculations are in good agreement with the experiments and further reveal the ballistic nature of cross-plane heat transport in PtSe$_2$ up to a certain thickness (~20 nm). Moreover, we show strong interlayer interactions in PtSe$_2$, short acoustic phonon lifetimes in the range of picoseconds, out-of-plane elastic constant $C_{33}$=31.8 GPa and a layer-dependent group velocity ranging from 1340 ms$^{-1}$ in bilayer PtSe$_2$ to 1873 ms$^{-1}$ in 8 layers of PtSe$_2$. The potential of tuning the lattice thermal conductivity of 2D layered materials with the level of crystallinity and the real-time observation of coherent phonon dynamics open a new playground for research in 2D thermoelectric devices and provide guidelines for thermal management in 2D electronics.

**1. Introduction**

Atomically thin 2D semiconductors have attracted immense attention in the scientific community due to their exceptional layer-dependent optical, electronic and thermal properties that open new prospects in the microelectronics industry.[1] Vertical devices consisting of one- or few-atom thick 2D materials, where heat transport usually occurs in the vertical direction, have already shown excellent performance in diodes,[2] photodetectors,[3] transistors[4,5] and solar cells.[6] In such devices, the interfacial thermal properties of atomically-thin layered 2D materials significantly vary depending on their thickness, interlayer interactions and degree of bonding with the substrate. It is therefore essential to understand cross-plane thermal transport and phonon dynamics, particularly in noble metal dichalcogenides, like PtSe$_2$, due to their large potential of integration in future high performance 2D devices.[7–10]

PtSe$_2$ exhibits outstanding inherent properties, including high room temperature carrier mobility, which is 8 times larger than MoS$_2$,[11] excellent stability to air and resistance to oxidation, better than black phosphorous.[5] Moreover PtSe$_2$ can be easily integrated in practical devices since it can be grown at low temperatures.[12] Together with its widely tunable bandgap[13] and layer-dependent semiconductor-to-semimetal transition behaviour,[14] it is considered as a promising material to be employed in many electronic,[5] optoelectronic[7] and thermoelectric devices.[15] For instance, Moon *et al.*[15] have recently showed that band engineering by thickness modulation leads to a 50-fold enhancement of the thermopower in bilayer PtSe$_2$ nanosheets with respect to bulk PtSe$_2$. Moreover, calculated results have shown that in monolayer PtSe$_2$ compressive or tensile strain can induce significantly enhanced n- or p-type Seebeck coefficients.[16] Despite the large potential of PtSe$_2$ in thermoelectric applications, experimental studies of the intrinsic thermal properties[17] and phonon dynamics of PtSe$_2$[18] films are few and still limited.

In this work, we study the phonon dynamics and thermal properties of supported crystalline and polycrystalline PtSe$_2$ thin films of varying thickness (0.6 – 24 nm), which were grown by molecular beam epitaxy (MBE) on zinc oxide (ZnO) substrates. First, by using a combination of low-frequency Raman and pump-probe coherent phonon spectroscopies we investigate the layer-breathing modes (LBM) in PtSe$_2$ thin films and extract an effective out-of-plane elastic constant, a layer-dependent sound velocity and the acoustic phonon lifetimes. Then, we focus on unravelling the impact of crystallinity and size effects on the cross-plane thermal conductivity of supported PtSe$_2$ thin films taking into account the interfacial thermal resistances in our multilayer sample geometry. Finally, we investigate the thermal conductivity by the first-principles Boltzmann transport equation (BTE) computing the harmonic and anharmonic force constants by density functional theory (DFT) to reveal the microscopic mechanism of heat transport in PtSe$_2$ thin films.

**2. Results and Discussion**

## 2.1. Material growth, structural characterization and phonon dynamics

Two PtSe$_2$ wedges were grown under ultrahigh vacuum (base pressure in the low 10$^{-10}$ mbar range) in an MBE chamber equipped with a cryo-panel and a Reflection High Energy Electron Diffraction (RHEED) setup. For the crystalline PtSe$_2$ wedge, four monolayers of PtSe$_2$ were deposited by co-evaporating Pt and Se on the ZnO (0001) substrate kept at 450°C. The resulting RHEED patterns were anisotropic with a 7° of mosaicity (see Fig. S1 in Supplementary), demonstrating the good-crystalline character of the film, and the epitaxy relationship was found to be ZnO (0001) [100]/PtSe2 (111) [100]. Additional characterization of epitaxial ZnO/PtSe$_2$ is reported in reference 11. For the poly-crystalline PtSe$_2$ wedge, we first deposited a 2.4 Å-thick Pt film by magnetron sputtering at room temperature and selenized it in the MBE chamber by deposition of Se at room temperature and subsequent annealing at 750°C under Se flux. The magnetron sputtering reactor and the MBE chamber are being connected under ultrahigh vacuum and PtSe$_2$ films are fully grown in situ. After selenization, the equivalent PtSe$_2$ thickness was two monolayers (2 ML). The RHEED patterns were streaky but isotropic, showing the in-plane poly-crystalline character of the film, with a well-defined (0001) surface for all grains but random in-plane crystal orientation. Two more PtSe$_2$ films were finally deposited by co-evaporating Pt and Se at 450°C to obtain a 4 ML thick poly-crystalline film (see Fig. S1c in Supplementary).

In a second step, we covered the sample with a motorized mechanical mask except a thin 1.5 mm-large band at the edge to deposit 14 ML of PtSe$_2$ by co-evaporating Pt and Se at 450°C. The mask was then retracted at constant speed and stopped at 1 mm from the edge to evaporate the 0-22 ML PtSe$_2$ wedge by co-evaporation at the same temperature. We checked by RHEED that the films retained their respective single and poly-crystalline character at the end of the growth. The final structure is shown in Fig. S1d of the Supplementary. Each PtSe$_2$ monolayer consists of three atomic sublayers, in which Pt atoms are sandwiched between Se atoms (see Fig. 1a). Both samples were annealed at 750°C for 10 minutes under Se flux after the growth in order to improve their crystalline quality. To avoid the film degradation during air transfer, the films were capped by a ~10 nm-thick amorphous Se layer deposited at room temperature.

The crystalline, acoustic and morphological characterizations of the samples were studied by Raman spectroscopy and high-resolution scanning transmission electron microscopy (HR-STEM) measurements. The arrangements of Pt and Se atoms in the planar HR-STEM image (Fig. 1b) and the symmetric diffraction pattern in the fast Fourier transform (FFT) image (Fig. 1c) further support the good crystal structure nature of the films. In addition, the high-magnification cross-sectional HR-STEM images (Fig. 1d, e) show the layered hexagonal honeycomb structure of a 5 layers $PtSe_2$ film, where bright Pt atoms are surrounded by six lighter colored Se atoms. The in-plane and out-of-plane lattice parameters were obtained from the ($l$ 0 0) and ($l$ $m$ 0) planes of the FFT and image analysis of cross-plane sections of the film (see Fig. 1e), respectively. The in-plane lattice constants were found to be $a = b \approx 0.37$ nm and $c \approx 0.52$ nm in good agreement with previous studies[10,15,19] and excellent agreement with the ones deduced from our x-ray diffraction measurements (see Fig. S2 in Supplementary). On the other hand, DFT calculations give $a = 0.377$ nm and $c = 0.486$ nm. Whereas $a$ is in excellent agreement with our measurements, $c$ is significantly underestimated (6.5%). There may be multiple reasons for this discrepancy, among which the difficulty in the DFT calculations to simulate $PtSe_2$ and PtSe compounds for all the GGA exchange functionals (e.g., PBEsol, AM05)[20] and even van der Waals functionals (e.g., vdW-DF-C09 and vdW-DF-CX).[21]

The layered $PtSe_2$ adopts a T-type hexagonal crystal structure, belonging to the $P\bar{3}m1$ space group with a $D_{3d}$ (-m) point group. The primitive $PtSe_2$ cell contains 3 atoms, then its vibrational spectrum includes 9 modes: 3 acoustic ($A_{2u} + E_u$) and 6 optical ($A_{1g} + E_g + 2A_{2u} + 2E_u$). The optical modes can be classified as Raman active ($E_g$, $A_{1g}$ and LO ($A_{2u} + E_u$)) and infrared active ($2E_u + 2A_u$) modes.[22] Figures 1f, g show the Raman spectra of single- and polycrystalline $PtSe_2$ films of different thicknesses, respectively. The high frequency modes (170-210 cm$^{-1}$) are originated from in-plane and out-of-plane vibrations of Se atoms corresponding to $E_g$ (~180 cm$^{-}$ [1]) and $A_{1g}$ (~205 cm$^{-1}$) modes, respectively. In addition, a weak interlayer longitudinal optical (LO) mode can be also observed at ~235 cm$^{-1}$. LO is generated by a combination of the in-plane $E_u$ and out-of-plane $A_{2u}$ modes from the vibrations of Pt and Se atoms in opposite phase. The Raman peaks located in the low frequency regions (-30 to 30 cm$^{-1}$) correspond to interlayer vibrations of PtSe$_2$ planes. [5,18]

Specifically, the vibrations detected here are related to the out-of-plane displacements of the PtSe$_2$ layers known as layer-breathing modes (LBM). As shown in Fig. 1f, g, the position of the LBM shifts to lower frequencies as the number of layer increases. The thickness dependence can be simulated by using a one-dimensional linear atomic chain model.[23] This model considers each PtSe$_2$ layer as a large atom with an effective mass per unit area $\mu = 4.8 \times 10^{-6}$ kg m$^{-2}$ connected by a string with an effective interlayer breathing force constant (IBFC) per unit area $K_\perp$ and separated by a distance $d \approx 0.52$ nm given by the interatomic distance between layers (see Fig. 1a). The solution of the linear chain with vanishing stress as boundary condition on the free surface is given by:

$$f = \sqrt{\frac{K_\perp}{\mu \pi^2}} \sin\left(\frac{q_{N,j}}{2} d\right) \quad (1)$$

where $f$ is the vibrational frequency of the mode, $q_{N,j} = 2\pi/\lambda_{N,j}$ is the acoustic wavevector, $\lambda_{N,j} = 2Nd/j$ is the phonon wavelength, $N$ is the number of layers ($N = 2, 3, …$) and $j$ is the index of the acoustic mode ($j = 1, 2, 3…$). The experimental and the fitted thickness-dependence of the first LBM ($j = 1$) is shown in Fig. 2a. The fitted curve was obtained using the IBFC adjustable variable, which is determined to be $K_\perp = 6.0 \pm 0.14 \times 10^{19}$ N m$^{-3}$. Similar value was also obtained by Chen et al[18] with $K_\perp = 6.2 \times 10^{19}$ N m$^{-3}$ using the same approach. Now, if we multiply the IBFC by the interlayer distance, we can also derive the corresponding $C_{33}$ component of the elastic constant tensor.[24,25]

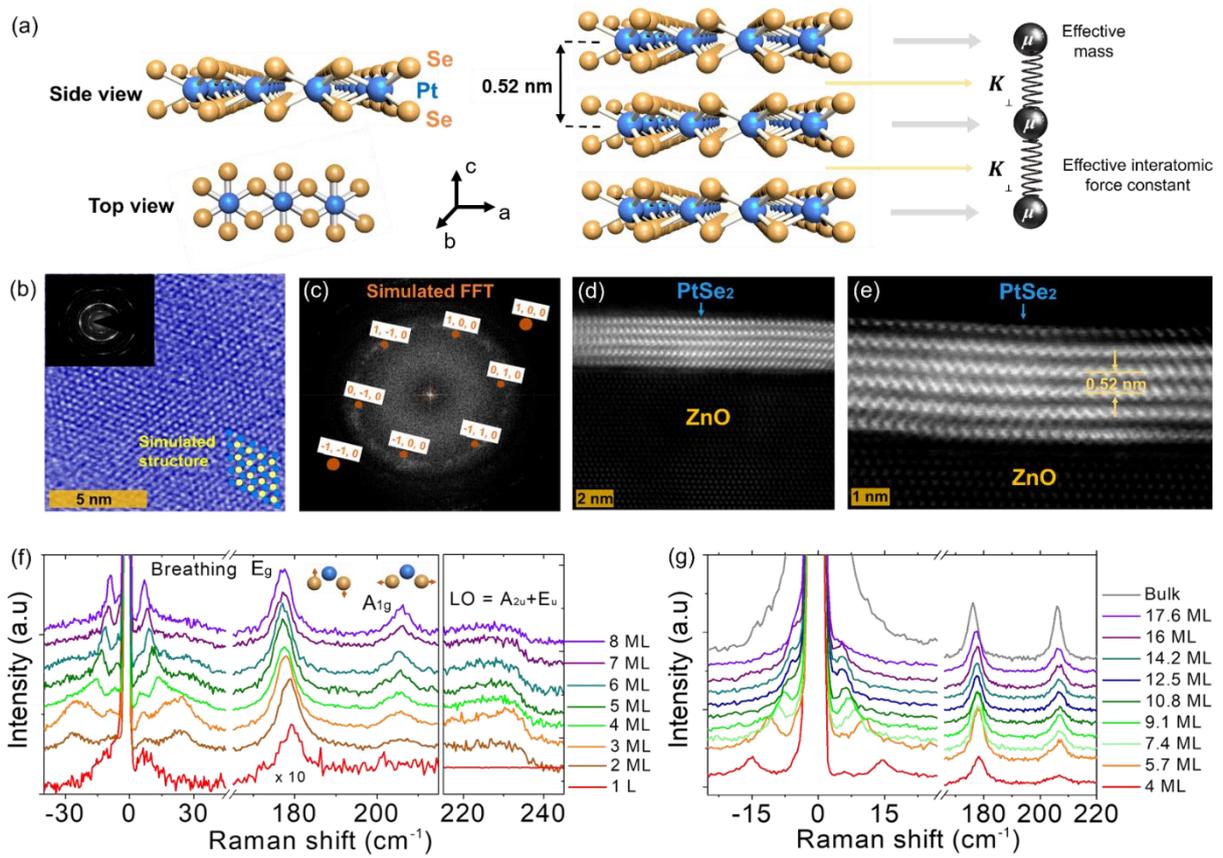

**Fig. 1 Structural characterization of PtSe$_2$ layers.** (a) Side and top views of the PtSe$_2$ crystal structure (left), schematics of the side view of a trilayer PtSe$_2$ crystal structure (middle) and the one-dimensional linear atomic chain model (right). The blue and orange spheres represent Pt and Se atoms, respectively. (b) HR-STEM images of a 5 ML single-crystalline PtSe$_2$ film and (c) the Fast Fourier transform of the respective image, from which we extract the *a* and *b* lattice constants, $a = b \approx 0.37$ nm. (d, e) Cross-sectional TEM images of the as grown 3 nm (5 ML) PtSe$_2$ film, from which we extract the *c*-lattice constant, $c \approx 0.52$ nm. The PtSe$_2$ is layered and each layer is parallel to the underlying ZnO substrate. Thickness-dependent Raman spectra of (f) crystalline and (g) polycrystalline PtSe$_2$ films deposited on ZnO. The inset in (f) shows blue and orange spheres that represent Pt and Se atoms, respectively, and arrows that point to the direction of the movement of that layer.

For our system we found $C_{33} = 31.8 \pm 0.95$ GPa, which is in the same range with other 2D materials such as, e.g., 38.7 -36.5 GPa for graphite, 54.3 GPa for MoSe$_2$[26], 24.5 GPa for h-BN[27], 52 GPa and 52.1 GPa for MoS$_2$ and WSe$_2$, respectively.[25] It is interesting to notice that the polycrystalline films also follow the same trend than the crystalline samples (see Fig. 2a) thus most likely the granular characteristics of the sample do not affect significantly the interlayers forces. Likewise, it is also possible to obtain the C$_{44}$ constant using the effective interlayer shear force constant ($K_{//}$). Chen et al.[18] calculated $K_{//}$ = 4.6 x 10$^{19}$ N m$^{-3}$ for PtSe$_2$ based on density functional theory simulations from Zhao,[5] giving C$_{44}$ = 23 GPa and $v_{TA}$ = 1873 m s$^{-1}$. In addition, from Eq. (1), it is also possible to extract the group velocity considering that $v_g = d(2\pi f)/dq$.

$$v_g = d\sqrt{\frac{K}{\mu}} \cos\left(\frac{\pi j}{2N}\right) \qquad (2)$$

In the limit N→∞ and considering the experimental interlayer distance $d \approx 0.52 \pm 0.01$ nm, Eq. (2) gives the bulk limit for the cross-plane longitudinal velocity $v_{LA}$ = 1873 ± 42 m s$^{-1}$. A similar value can be obtained using an approximation of $v_g = \sqrt{\frac{C_{33}}{\rho}}$ =1825 ± 28 m s$^{-1}$ where $\rho$ = 9540 kg m$^{-3}$ is the density of PtSe$_2$ and C$_{33}$ = 31.8 GPa. Using a numerical differentiation, it is also possible to extract the group velocity as a function of the number of layers as shown in Fig. 2b.

Furthermore, the phonon dynamics of the crystalline PtSe$_2$ samples were measured by pump-probe coherent phonon spectroscopy using asynchronous optical sampling method (ASOPS) (see Fig. 2c and d). In this method two pulsed lasers (pump and probe beams) are focused on the sample surface. The pump produces a change of reflectivity which is measured by the probe beam as a function of the time delay between the lasers. The absorption of the pump laser causes an increase of the local strain via two separate mechanisms, namely thermal expansion, and the hydrostatic deformation potential. The thermal expansion is a consequence of the anharmonicity of the lattice, whereas the deformation potential is due to the excitation of electrons into binding orbitals. Both mechanisms periodically change the effective volume of the material, which in turn modulates the optical properties of the film that are probed by the second laser.

In this experiment, we used 2 fs Ti:sapphire lasers with a repetition rate of ~ 1 GHz stabilized via an electronic feedback loop to achieve a small repetition rate difference of 2 kHz. The rate difference produces a time delay between the pump and probe pulses without the need of mechanical delay stage.[28,29] The delay between the pump pulses is kept at 1 ns, while the pump and probe pulses coincide every 500 ms. The measurements were done at a fixed central wavelength of λ = 790 nm for the pump (~1 mW) and the probe (~0.15 mW) beams, collinearly focused to a ~ 1 μm-diameter spot on the sample surface. Figure 2c shows the typical modulation of the reflectivity (ΔR/R) as a function of the time delay (black open circles) and the corresponding best model fit in a 5 ML PtSe$_2$ single crystalline film. The change of reflectivity includes the effects of a fast and slow electronic relaxation processes as well as the dynamics of the generated phonons. The time traces are fitted to damped harmonic oscillator of the form:[18]

$$\Delta R/R = \sum_{i=1}^{2} A_i \exp\left(\frac{-t}{\tau_i}\right) + \sum_{j=1}^{2} B_j \exp\left(\frac{-t}{\tau_j}\right) \sin(2\pi f_j t + \phi_j) \qquad (3)$$

where $t$ is the time delay between the pump and the probe lasers, A and B are the amplitudes, $\tau_i$ are the decay times, $f$ the phonon frequency and $\phi$ is a phase delay. The first term of Eq. (3) represents the fast and slow relaxation processes of the excited carriers. The second term describes the damped phonon oscillations. The Fast Fourier transforms (FFT) of the time-domain data for of all measured samples are displayed in the Supplementary.

Figure 2d shows the measured decay time of LBM as a function of the number of layers (yellow dots) and the calculated phonon decay due to boundary scattering contribution based on the model proposed by Ziman.[30] This model considers the impact of the roughness ($\eta$) of the film surface through a phenomenological parameter $p = \exp[-16\pi^2\eta^2/\lambda^2]$, which modifies the effective mean free path $\Lambda = (1+p)/(1-p) \, Nd/v_g$ of the acoustic waves. The black and red curves of Fig. 2d were calculated using a thickness dependent group velocity (Eq. (2)) and a constant value given by the bulk limit (1873 m s$^{-1}$), respectively, with a constant $\eta = 0.82$ nm as measured by AFM (see Fig. S3 in Supplementary). Apart from the good agreement between the theory and the experimental data, we observe that the phonon lifetime is not affected by the group velocity (thickness dependent or constant). Similar observations of boundary-limited lifetimes have been reported in suspended MoSe$_2$ flakes by Soubelet et al.[26] They measured lifetimes of the order of ~1-41 ps in 2-8 ML limited by boundary effects. While in thicker samples (> 20 ML), they observed longer lifetimes (~0.3-10 ns) limited by phonon-phonon scattering. The large difference in lifetimes measured here (1-8 ps) is mainly attributed to the difference in the phonon group velocity which is 1.4 to 1.5 times smaller than MoSe$_2$.

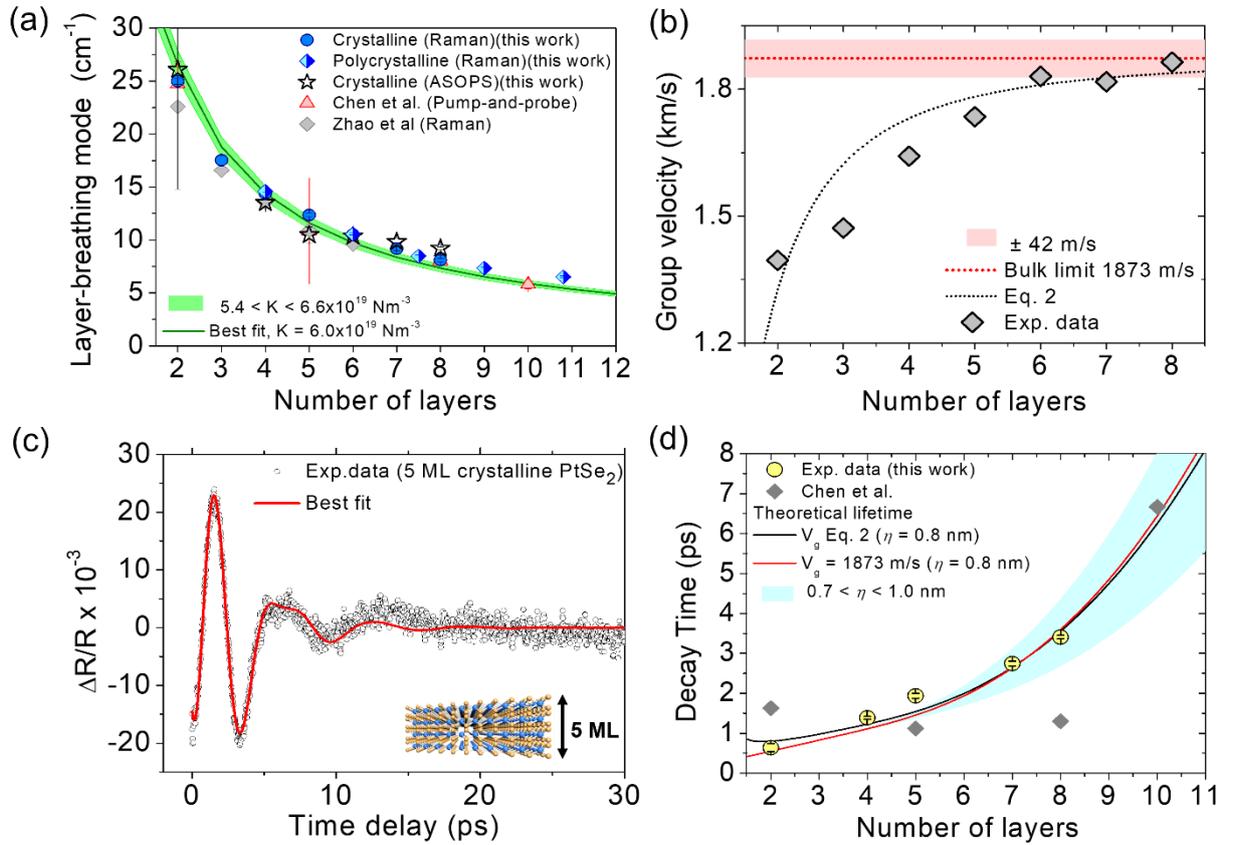

**Fig. 2 Phonon dynamics in PtSe$_2$.** (a) Peak position of the first order breathing mode measured in crystalline and polycrystalline PtSe$_2$ films using low-frequency Raman spectroscopy and ASOPS as a function of the number of layers and comparison with other experimental works. (b) Calculated group velocity as a function of the number of layers. (c) Representative transmission signal obtained with ASOPS in 5 ML PtSe$_2$ (black open circles) and the best fit (red line). (d) The extracted phonon decay time (yellow dots) vs the number of layers. The red and black curves are theoretical predictions of the phonon lifetimes based on the model proposed by Ziman.[30] The region displayed with light blue shows the impact of roughness in the calculated lifetimes. The grey data points show phonon lifetimes measured by Chen et al.[18]

## 2.2 Thermal conductivity and interfacial heat transport measurements

For the thermal measurements, we used a custom-built frequency domain thermo-reflectance (FDTR) setup, a well-established optical pump-probe technique, capable of measuring heat transport in thin films and across interfaces.[31–36] A schematic of the experimental setup is shown in Fig. 3a. A modulated pump beam with a wavelength of 448 nm is focused onto the sample, creating a periodic heat flux with a Gaussian spatial distribution on the sample surface. A reflected probe beam with a wavelength of 532 nm is aligned coaxially with the pump beam and focused with the pump spot to monitor the periodic fluctuations in reflectivity at the sample surface caused by the oscillating sample temperature (see Fig. 3b). All the samples after the MBE growth were coated with a 100 nm thick Au layer, which was chosen to maximize the coefficient of thermo-reflectance at the probe wavelength ($\sim 2.36 \times 10^{-4}$ K$^{-1}$). We used a lock-in amplifier (Zurich Instruments HF2LI) to record the amplitude and phase response of the reflected probe beam to the thermal wave, and the phase lag between the pump and probe beam as the observable quantity. More details on the experimental setup can be found elsewhere.[35]

In the present experiments, we obtained frequency-domain measurements by varying the modulation frequency of the pump beam over a wide range (20 kHz–40 MHz). The cross-plane thermal conductivity ($k_z$) of the PtSe$_2$ thin films was subsequently extracted by following a multilayer three-dimensional (3D) heat diffusion model that includes the interface thermal conductance between the different layers and anisotropic heat transport.[31] For each experiment, we quantified first the sensitivity of the recorded phase signal to different parameters: $k_z$, the volumetric heat capacity ($C$), the thermal conductivity anisotropy ratio ($k_r/k_z$), and the different interface thermal conductance's, in a similar manner to that of Schmidt et al.[31] The details about the measurement sensitivity to different combinations of parameters in the model can be found in the Supplementary (Fig. S4). In all the experiments we used a high-magnification 50× objective lens that produced a focused root-mean-square (rms) spot size (1/e$^2$ radius) of approximately ~1.5 μm. To reduce the uncertainties associated with variations in the laser spot size between measurements on different films, the spot size was measured for each different experiment using the knife's edge method, as shown in the SI (see Fig. S6). All the measurements were performed under both ambient and vacuum conditions, at room temperature ($T_{amb}$=22°$C$).

Before the thermal measurements on thin PtSe$_2$ films, we performed FDTR measurements on a bulk PtSe$_2$ crystal coated with 100 nm of Au in order to extract the anisotropy ratio of the thermal conductivities. The thermal conductivity of the deposited Au film was first measured using electrical conductivity and the Wiedemann−Franz law, $k_{Au}$ =195 Wm$^{-1}$K$^{-1}$ (see details in Supplementary, Fig. S7). The volumetric specific heat of Au and PtSe$_2$ were taken from literature.[10,36] This leaves us with three unknowns: the cross-plane and in-plane thermal conductivity of the bulk PtSe$_2$ crystal and the Au-PtSe$_2$ interface thermal conductance ($G_{Au\text{-}bulk}$). The in-plane thermal conductivity was expressed in terms of the anisotropic ratio ($\alpha$) and the cross-plane thermal conductivity $k_r = \alpha \cdot k_z$.

However, from the sensitivity analysis we found that the recorded phase signal has low sensitivity to $G_{\text{Au-bulk}}$ and only at high frequencies while its sensitivity to the anisotropic heat flow is high in almost the whole frequency range (see Fig. S5 in Supplementary). This allowed us to extract α directly from the model fit of the experimental data in the low frequency range (20 kHz–1 MHz) using a nonlinear least-squares routine, which requires an initial guess to determine the value of the free parameter ($k_z$) (see Fig. 3c). The inset in Fig. 3c shows the numerical errors of the applied model fits as a function of different anisotropy ratios. The minimum value of the fitting error, which gives us the best fit of the FDTR data, corresponds to α = 9, a cross-plane thermal conductivity of 4.6 ± 0.7 Wm$^{-1}$K$^{-1}$ and $k_r$ ≈ 41.4 Wm$^{-1}$K$^{-1}$. The observed anisotropy ratio between $k_r$ and $k_z$ in PtSe$_2$ is similar with previous reported values of other 2D-layered materials, such as InSe (α ~ 10),[37] SnSe$_2$ (α ~ 7-8)[35,38] and armchair black phosphorous (α ~ 6-10).[39] Our first-principles BTE calculations give $k_z$ = 2.84 Wm$^{-1}$K$^{-1}$ and $k_r$ = 39.0 Wm$^{-1}$K$^{-1}$. $k_r$ is in excellent agreement with the measurement, while $k_z$ is significantly overestimated, possibly due to the discrepancy in the equilibrium $c$ lattice parameter mentioned above. Nevertheless, we can still use DFT and BTE to analyse the phonon contributions to cross-plane transport in PtSe$_2$ bulk and the effect of finite thickness in thin films, at least qualitatively.

To study the impact of crystallinity and film thickness on the cross-plane thermal conductivity of PtSe$_2$ thin films, we performed FDTR measurements in large-area crystalline and polycrystalline PtSe$_2$ films of different thicknesses (0.6 – 24 nm) on ZnO substrate. The wedge samples stacks consist of Au/PtSe$_2$/ZnO (see Fig. 3b and Fig. S1 in Supplementary). Typical examples of the recorded phase signals and the corresponding best model fits for monolayer and 40 ML crystalline PtSe$_2$ are shown in Fig. 3d. Here, the key parameters for the model in our multilayer system are: the spot sizes of the pump and probe beams, $k_z$, the film thickness $t$, the $C$ of each layer, and the interface thermal conductances (between Au and PtSe$_2$ ($G_1$) and between PtSe$_2$ and ZnO ($G_2$)). The thermal conductivity of ZnO was measured with the 3ω technique,[40,41] $k_{ZnO}$= 55 Wm$^{-1}$K$^{-1}$ (see Fig. S7 in Supplementary). The spot sizes for each experiment were measured with the knife's edge method. The thickness of Au and PtSe$_2$ films were measured by AFM and low-frequency Raman spectroscopy, respectively (see section above and Supplementary). The heat capacities of Au, ZnO and PtSe$_2$ were taken from the literature.[10,36,42] Thus, the remaining unknown parameters were the $k_z$ value of the PtSe$_2$ films, $G_1$ and $G_2$.

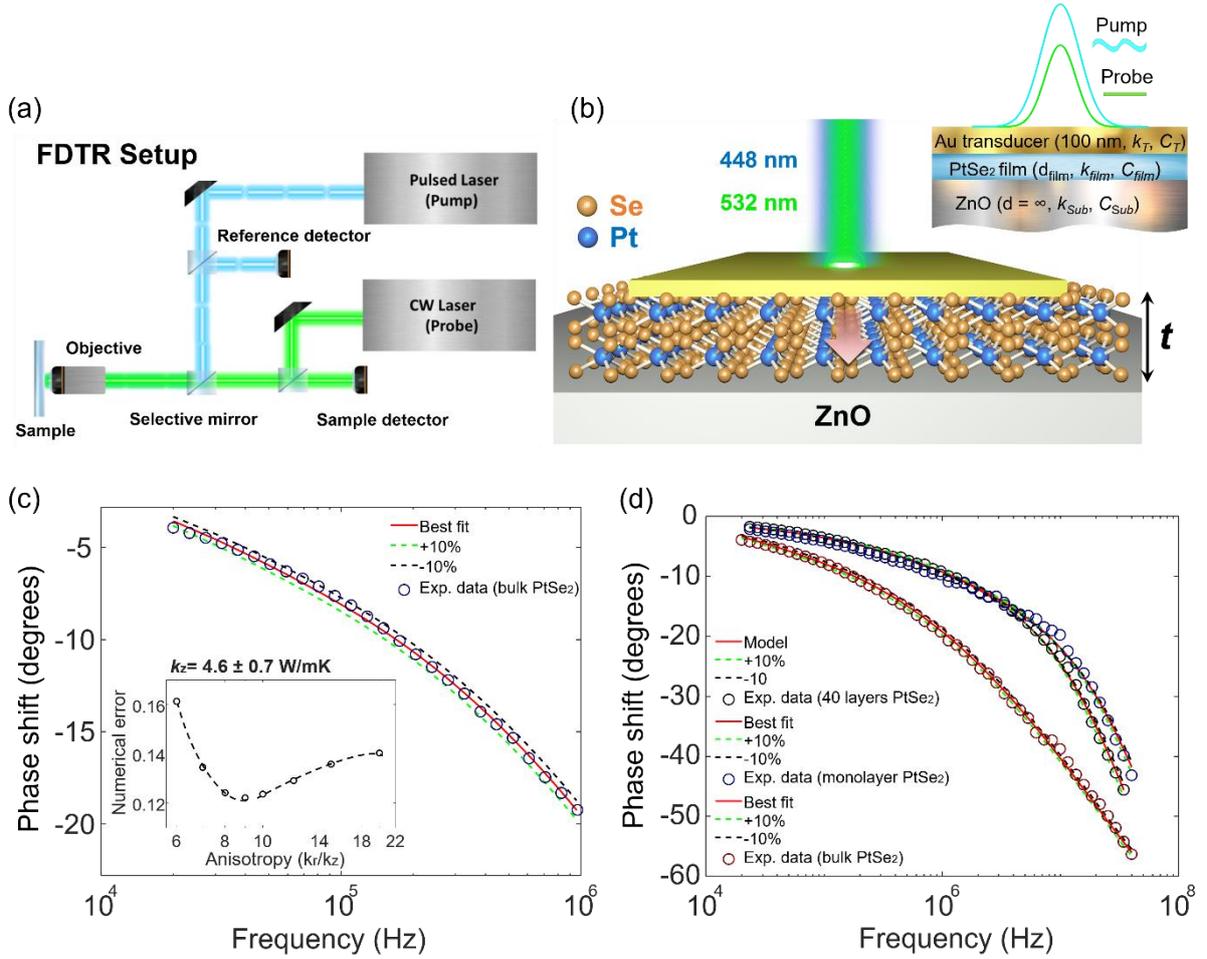

**Fig. 3 Frequency-domain thermoreflectance measurements.** Schematic illustrations of (a) the FDTR technique and (b) the multilayer system. The laser pump is directly modulated by the reference output signal from a lock-in amplifier while the lock-in detects the phase lag in the probe signal at the modulation frequency relative to the reference output signal. (c) FDTR data from bulk PtSe$_2$ crystal and the best model fit in a low frequency range (20 kHz - 1 MHz) for a ratio $k_r/k_z$ = 9. The inset graph shows the calculated numerical errors of fits with different

anisotropies. (d) Typical FDTR data measured in bulk PtSe$_2$ (red circles), 40 layers (black circles) and monolayer (blue circles) crystalline PtSe$_2$ films and the corresponding best model fits in the whole frequency range.

To extract a unique value of $k_z$ from a single measurement we followed a fitting approach similar to that suggested in previous works[31,32,35] and supported by our sensitivity analysis (see Fig. S4 in Supplementary). First, we estimate $k_z$ by fitting to experimental data in a low frequency range (20 kHz - 1 MHz), where the measurement sensitivity to the $G_1$, $G_2$ and heat capacity of the films is negligible. Then, we fix $k_z$ and fit to experimental data in a high frequency range (1 MHz - 40 MHz) to estimate simultaneously $G_1$ and $G_2$. As initial guess to determine the value of the free parameters ($k_z$, $G_1$ and $G_2$) we used previous reported values of similar material systems.[43–45] We also verified that the final fit results are not sensitive to the choice of initial values. The same analysis has been followed to extract $k_z$, $G_1$ and $G_2$ values for all the PtSe$_2$ films.

It is interesting to note that, in contrast to the bulk case, in FDTR measurements in thin films the diameter of the laser spot is usually large compared to the thermal diffusion length during the modulation period of the pump beam. Thus, the heat flow is expected to be mainly one-dimensional in the cross-plane direction.[32] This has been confirmed by our sensitivity analysis, where we found that the sensitivity of the recorded phase signal to the in-plane transport is relatively low (see Fig. S4c in Supplementary). Therefore, since anisotropic differences in thermal conductivities as a function of the film thickness cannot be resolved, for the data analysis we used the anisotropy ratio extracted from the bulk PtSe$_2$ experiments. The dependence of the thermal anisotropy with the film thickness has been studied recently in similar material systems, such as supported InSe[37] and SnSe$_2$[35] films, showing a thickness-independent $\alpha$ of 10 and 8.4, respectively.

Figure 4a displays the extracted $k_z$ value of all the PtSe$_2$ thin films as a function of thickness. In both crystalline (black data points) and polycrystalline films (green data points), we observe a linear increase of $k_z$ with increasing the film thickness starting from 2.4 nm (~ 4 monolayers) up to 10 nm and 24 nm, respectively. In polycrystalline films with thicknesses $t$ >12 nm, $k_z$ shows a plateau and a maximum cross-plane thermal conductivity reduction of approximately 35% was observed as compared to the crystalline samples. This result highlights the strong impact of crystallinity on $k_z$ and suggests that depending on the film crystallinity phonons with different mean free paths (MFPs) contribute to the cross-plane thermal conductivity. In crystalline PtSe$_2$ films, heat is propagating through coherent vibrations (phonon modes) that travel distances at least 24 nm (~40 monolayers) while in polycrystalline samples, they start to decay above 12 nm (~20 monolayers). This behavior most likely is attributed to the enhanced phonon scattering due to the presence of a high density of defective grain boundaries randomly oriented in the polycrystalline film.

In Fig. 4a, we also show the $k_z$ obtained from first-principles DFT-BTE calculations, both, for bulk PtSe$_2$ and thickness dependent values as two different sets of data. The first set accounts solely for the finite thickness of the film and gives an overall much larger thermal conductivity than the measured one at the corresponding thickness (red data points). The overestimation of the calculated thermal conductivity for the whole thickness range can be related to the underestimation of the *c*-axis compared with the experimental data and the assumption of a perfect single-crystal structure of the films. In addition, $G_1$ and $G_2$ were not taken into account in the DFT calculations, which further explains the overestimated values in the calculated thickness dependence of $k_z$. However, we observe that the trend of thickness-dependent of the thermal conductivity is preserved. Finally, correcting the BTE calculation for the observed thickness-dependent reduction of sound velocity provides better agreement between theory and experiments over the whole range of thickness considered (yellow-black diamonds).

The observed linear dependence of $k_z$ from 4 monolayers up to a certain thickness in both sets of samples is consistent with ballistic heat conduction processes, as has been shown in previous works, where the thermal conductivity increased linearly with the characteristic length of the system.[43,46,47] However, in few-layer crystalline PtSe$_2$ films ($t$ < 2-3 nm), we observe a deviation from the linear thickness-dependence of $k_z$ (see Fig. 4a, black spheres). Since interfaces might dominate cross-plane heat transport in very thin films, the apparent deviation might be attributed to variations in the total thermal resistance per unit area, $R_{tot}$, of the film (see calculations below), which can be written as the sum of the combined interface thermal resistance, $R_{int}$= $1/G_1$ + $1/G_2$, and volumetric cross-plane thermal resistance, $R_{PtSe2}$= $t/k_z$.[43,48] Note that this expression is not valid in very thin films and $G_1$, $G_2$ should be treated as one diffusive interface instead of two discrete ones.[43,48] Another possible explanation for this deviation is the semiconductor-to-semimetal evolution of PtSe$_2$ films after 4 monolayers (~ 2 nm),[49] which might result in an additional contribution of the electron thermal conductivity to $k_z$ and/or lower contact resistivity between PtSe$_2$ and Au.[12]

To quantify the impact of cross-plane ballistic phonon transport on the total thermal resistance of multilayer PtSe$_2$ films, we plotted the total thermal resistance, $R_{tot}$=$R_{PtSe2}$+$R_{int}$, versus film thickness (Fig. 3b). We observe that in crystalline PtSe$_2$ $R_{tot}$ remains almost constant in the entire thickness range (see black spheres), with values approximately of $28 \pm 10$ m$^2$KGW$^{-1}$, while in polycrystalline samples (see green spheres) $R_{tot}$ is constant only up to 12 nm and then starts to increase up to $75 \pm 10$ m$^2$KGW$^{-1}$. The large contribution of the volumetric resistance component, $R_{PtSe2}$ to $R_{tot}$ and the thickness-independent $R_{tot}$ in crystalline films indicate ballistic phonon transport along the $c$-axis of the films, as has been previously observed in few-layer-graphene and single crystal MoS$_2$ flakes that showed a thickness-independent $R_{tot}$ for film thicknesses between 0.4 - 4 nm[48] and 20 - 40 nm,[43] respectively. The increased values of $R_{tot}$ in polycrystalline films after 12 nm indicate a transition from ballistic to diffusive transport regime most likely due to phonon grain boundary scattering. For comparison, in Fig. 4b we plot our calculated $R_{tot}$ (black and green spheres) with previous total cross-plane thermal resistance measurements on different 2D materials, where similar interfacial contributions from bottom (2D material/substrate) and top (metal/2D material) interfaces were taken into account.[43,48] Our values are in line with these estimations and suggest that in sufficiently thin films, phonons (especially long-wavelength phonons) can directly propagate between the metal and substrate without being strongly scattered by interfaces.

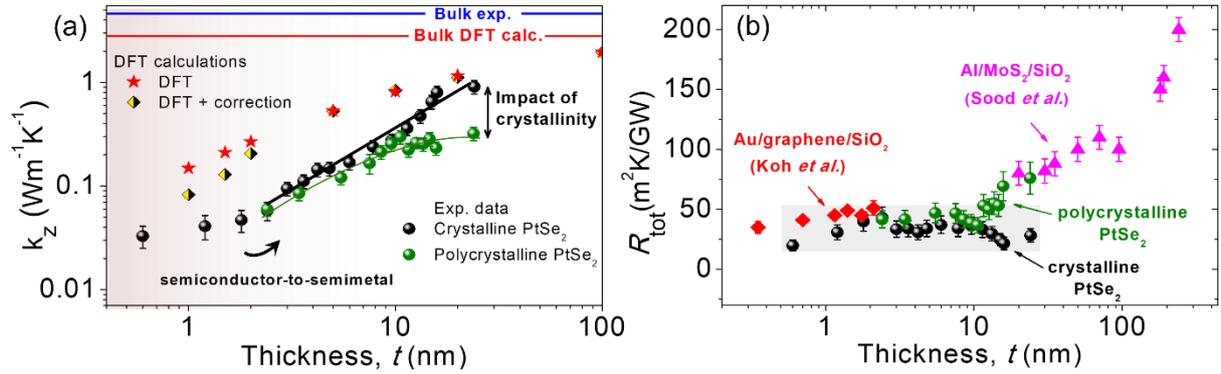

**Fig. 4 Thermal conductivity and interfacial heat transport measurements**. (a) Cross-plane thermal conductivity $k_z$ of crystalline (black spheres) and polycrystalline (green spheres) PtSe$_2$ films versus film thickness $t$. The uncertainty of the estimated $k_z$ was calculated based on error propagation for the input parameters. The blue and red lines in (a) display the bulk cross-plane thermal conductivity values obtained by FDTR and DFT calculations, respectively. The black and green curved are a guide for the eye. (b) Total thermal resistance, $R_{total} = R_{int} + R_{PtSe2}$, of crystalline (black spheres) and polycrystalline (green spheres) PtSe$_2$ thin films, plotted versus film thickness $t$. The uncertainty region is displayed as a grey rectangle. The total cross-plane thermal resistance measurements of Au/graphene/SiO$_2$[48] (red diamonds) and Al/MoS$_2$/SiO$_2$[43] (purple triangles) interfaces are also reported.

To further confirm the robustness of our approach to measure the intrinsic cross-plane thermal conductivity of thin films, we performed Raman thermometry measurements in few layers crystalline PtSe$_2$ films (2, 3 and 4 layers of PtSe$_2$, see Fig. S8 in Supplementary), where we found $k_z$ values in good agreement with the FDTR results (see Table S1 in Supplementary). The agreement between Raman and FDTR experiments further supports our $R_{tot}$ calculations and show that it is possible to quantify the interfacial thermal contributions in a multilayer structure from a single set of FDTR measurements despite the low phase sensitivity to $G_1$ and $G_2$. Note that in Raman thermometry measurements we have only the contribution of $R_2$ to the total thermal resistance.

As discussed above, to calculate the frequency response of the surface temperature to the pump beam we followed a multilayer 3D heat diffusion model that considers anisotropic transport.[33] To validate if a diffusive model is suitable, we estimate the room temperature phonon MFP in the cross-plane direction ($\Lambda_z$) using a simplified expression from the kinetic theory, $k_z \sim (1/3)Cv_s\Lambda_z$, where $v_s$ is the averaged sound velocity of cross-plane acoustic modes. For PtSe$_2$ we used C $\sim$ 1.78 MJm$^{-3}$K$^{-1}$ [50] while $k_z$ and $v_s$ were determined experimentally ($k_z \sim$ 4.6 Wm$^{-1}$K$^{-1}$, $v_s \sim \sqrt[3]{\frac{1}{3}(\frac{1}{v_{LA}^3} + \frac{2}{v_{TA}^3})}$) through our FDTR experiments and ASOPS measurements described above. From this estimation, we found $\Lambda_z \sim$ 5 nm, which corresponds to a thickness of approximately 10 layers. However, our FDTR experimental data suggest that in crystalline PtSe$_2$ films phonons with $\Lambda_z$> 24 nm substantially contribute to the thermal conductivity (about 37% of the bulk value, from our BTE-DFT calculations) while in polycrystalline films grain boundary scattering reduces this value to approximately 12 nm (see Fig. 4a). This is in close agreement with our DFT calculations that show that phonons with MFP larger than 15 nm contribute to almost 60% of the total cross-plane bulk thermal conductivity (see Fig. 5d).

Moreover, DFT-BTE calculations provide further insight in the heat transport mechanisms in PtSe$_2$. Figure 5a shows the phonon dispersions of bulk PtSe$_2$ along of high-symmetry paths in the Brillouin Zone. As opposed to 2H phases, we observe no gap between the optical and the acoustic modes. It is worth noting that the high-frequency optical modes are unusually dispersive in the Γ-A direction, if compared to other layered materials, for which optical phonons dispersions are usually flat.[51] This can be seen in Fig. 5b which displays the group velocities along the same k-point paths. Figure 5b also shows that the speed of sound in the Γ-A direction, averaged over the three acoustic branches, is 1930 m s$^{-1}$ ($v_{LA}$ =2120 m s$^{-1}$). In addition, in Fig. 5c we show that the MFPs in the cross-plane direction of the acoustic modes may be as high as 200 nm. Remarkably, there is a large number of optical modes with MFPs between 1 to 10 nm. This leads to a significant contribution of the optical modes to the thermal conductivity of bulk and films of PtSe$_2$, as shown in Fig. 5d and Fig. S10 in the SI.

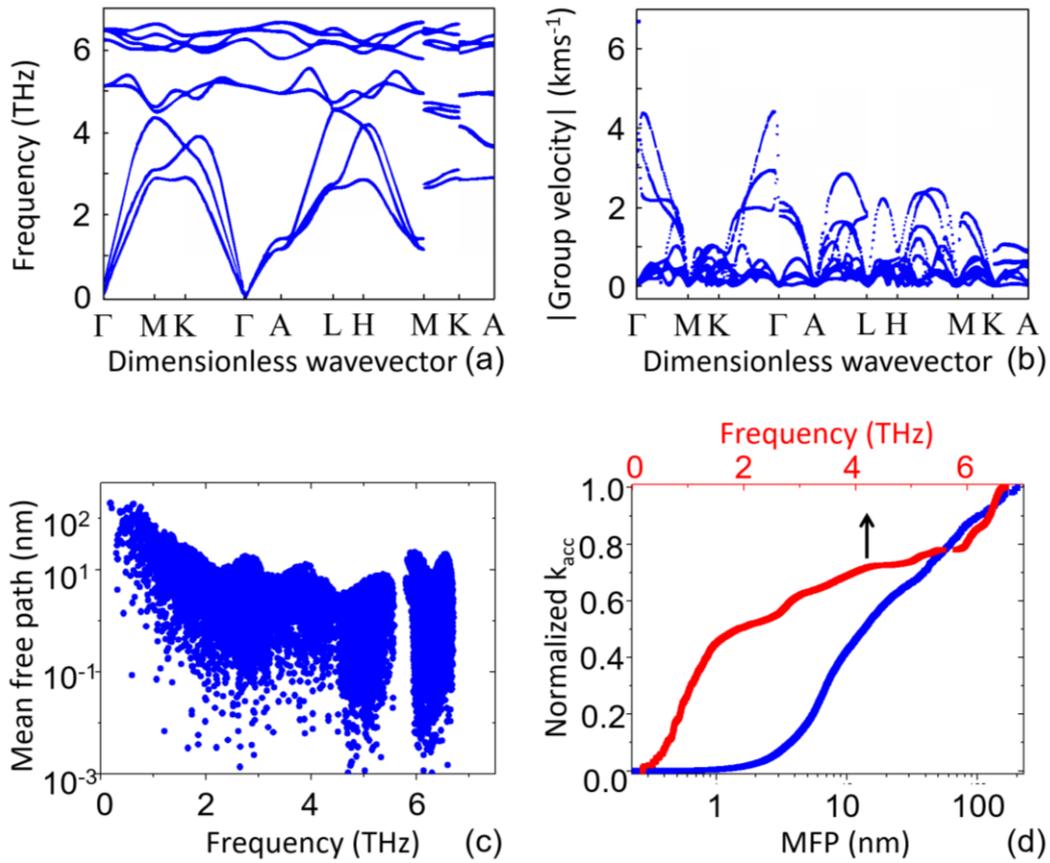

**Fig. 5 First-principles DFT calculations.** (a) Phonon dispersion relations and (b) group velocity of bulk PtSe$_2$ along high-symmetry directions in the Brillouin zone. (c) Calculated bulk phonon mean free as a function of frequency. (d) Normalized thermal conductivity accumulation ($k_{acc}$) as a function of the phonon frequency (red curve) and MFP (blue curve) for bulk PtSe$_2$.

In particular, Fig. 5d shows the normalized cumulative thermal conductivity as a function of frequency (red axis) and phonon mean free path (black axis) for bulk PtSe$_2$. We observe a quite narrow MFP distribution (1-200 nm) but much longer than the rough estimation from the kinetic theory ~5 nm. This discrepancy it is well known and comes from the fact that each phonon contributes in a different manner to the thermal conductivity.[43,52–54] An averaged value can be very misleading and underestimate the real contribution of the phonon MFP to the total thermal conductivity. Finally, it is interesting to note the rather large contribution (~ 30%) of low-frequency phonons (< 1 THz) to the total thermal conductivity. These phonons modes are sensitive to the introduction of additional periodicity and their dispersion relation can be easily tuned through nanofabrication.[55,56] This opens the possibility of exploring alternative ways of tuning thermal conduction through phonon engineering in two-dimensional materials.

## 3. Conclusion

In conclusion, we have studied phonon dynamic properties and heat transport in MBE-grown crystalline and polycrystalline PtSe$_2$ thin films of varying thickness (0.6 – 24) nm using a combination of characterization techniques, i.e., FDTR, low-frequency Raman and pump-probe coherent phonon spectroscopies, and state-of-the-art DFT calculations. Our work demonstrates the ability to quantify the influence of thickness and crystallinity on the cross-plane heat propagation in thin layered PtSe$_2$ films, showing an almost 35% reduction in the thermal conductivity of polycrystalline films with thickness larger than 12 nm in comparison with crystalline films of the same thickness. Moreover, from the phonon dynamic study in crystalline PtSe$_2$ we extract an out-of-plane elastic constant $C_{33}$ =31.8 GPa and a layer-dependent group velocity ranging from 1340 m s$^{-1}$ in bilayer PtSe$_2$ to 1873 m s$^{-1}$ in 8 layers of PtSe$_2$. Last, we showed that acoustic phonons in PtSe$_2$ thin films, which are the main carriers of heat in semiconductors, have extraordinarily short lifetimes in the order of picoseconds. Our results provide a new insight into the heat transport and phonon dynamics in 2D materials at the nanoscale, with potential implications for future design of 2D-based devices for energy harvesting and effective heat dissipation in thermoelectric and optoelectronic devices.

**Methods**

**MBE growth:** Pt and Se were respectively evaporated thanks to an e-gun evaporator and a standard effusion cell. The Pt deposition rate was set to 0.75 Åmin$^{-1}$ and monitored in real time by a quartz crystal microbalance. The pressure of Se at the sample position was measured with a retractable ionization gauge and set to $1.0\times10^{-6}$ mbar, which corresponds to a Se:Pt flux ratio of about 15. The MBE reactor is connected under UHV to a magnetron sputtering chamber with a Pt target. The Pt deposition rate was set to 4.7 Åmin$^{-1}$ thanks to a retractable quartz crystal microbalance, the DC magnetron power was 1 W and the argon pressure $1.2\times10^{-2}$ mbar. The ZnO(0001) substrates (CrysTec GmbH) were first etched for 30 s with an HCl 1.8% solution and rinsed with deionized water. This chemical treatment was followed by annealing for 1h at 900°C in O$_2$ atmosphere, then annealing in UHV for 5 min at ~800°C.

**XRD:** Grazing-incidence x-ray diffraction measurements were performed with a SmartLab Rigaku diffractometer. The source is a rotating anode beam tube (Cu K$\alpha$ = 1.54 Å) operating at 45 kV and 200 mA. The diffractometer is equipped with a parabolic multilayer mirror and in-plane collimators of 0.5° on both source and detector sides defining the angular resolution. A K$\beta$ filter on the detector side eliminates parasitic radiations.

**DFT:** We computed the lattice thermal conductivity of 1T (octahedral) PtSe$_2$ by density functional theory (DFT) and the Boltzmann transport equation (BTE). In the DFT calculations, we used a van der Waals density functional with consistent exchange (vdW-DF-CX) that reliably predicts the structural and vibrational properties of several 2D materials, including transition metal dichalcogenides.[57,58] Valence Kohn-Sham wavefunctions are expanded on plane-waves basis set with a cutoff of 40 Ry, and projector augmented pseudopotentials are used to model screened nuclei.[59] A uniform 8x8x6 mesh of k-points was used to integrate the first Brillouin Zone. Phonon dispersion relations were computed by density functional perturbation theory interpolating a uniform 4x4x3 q-point mesh.[60,61]

**BTE:** Second-order and third-order force constants for the thermal conductivity calculations were computed by fitting the forces of 80 configurations of the 4x4x3 $PtSe_2$ supercell, in which the atoms are randomly displaced with a standard deviation of 0.01 Å using the hiPhive code.[62] Second-order and third-order force constants were cut off at 6.9 Å and 5.13 Å, respectively. Fitting employed recursive feature elimination with a limit of 300 features (out of 961). The fit gives a root mean square error on the forces of 0.0044 eV/A with $R^2 = 0.999$. The solution of the linearized phonon BTE was computed by directly inverting the scattering tensor on a 21x21x17 mesh of q-points, as implemented in kALDo.[63] The effect of the finite thickness of the samples is taken into account by including the boundary conditions in the scattering term as proposed by Maassen and Lundstrom.[64]

**Data availability**

All relevant data are available from the corresponding author on request.

**Acknowledgements**

This work has been supported by the Severo Ochoa program, the Spanish Research Agency (AEI, grant no. SEV-2017-0706) and the CERCA Programme/Generalitat de Catalunya. The authors acknowledge support from Spanish MICINN project SIP (PGC2018-101743-B-I00), and the EU project NANOPOLY (GA 289061). The LANEF framework (ANR-10-LABX-51-01) is acknowledged for its support with mutualized infrastructure. P.X. acknowledges support by Ph.D. fellowship from the EU Marie Sklodowska-Curie COFUND PREBIST (Grant Agreement 754558). A.E.S. acknowledges support by the H2020-MSCA-IF project THERMIC-GA No. 101029727.

**Author Contributions**

A.E.S. and E.C.A. conceived the project. A.E.S and E.C.A. built the FDTR setup, performed the thermal measurements and data analysis. A.E.S. performed the AFM measurements. E.C.A. and P.X. performed the Raman measurements. E.C.A. and A.G. performed the ASOPS measurements and analysis. F. B., A.M., C.V, H.B. and M. J. fabricated the MBE samples and performed the XRD measurements. P.X., M.S., A.E.S. and E.C.A. performed the TEM measurements and the structural analysis. D. D. and Z.C. performed the DFT calculations and provide support to the theoretical analysis. All authors reviewed and edited the manuscript and have given approval to the final version of the manuscript. The manuscript was written by A.E.S. and E.C.A. A.E.S. and P.X. contributed equally.

**Competing interests**

The authors declare no competing interests.

**Additional information**

Supplementary information. The online version contains supplementary material available. Reflection high energy electron and X-ray diffraction measurements, FDTR sensitivity analysis, spot size measurements, AFM measurements, three omega measurements, Raman thermometry measurements, ASOPS measurements, DFT calculations.

# Supplementary Information

# Phonon dynamics and thermal conductivity of PtSe$_2$ thin films: Impact of crystallinity and film thickness on heat dissipation

*Alexandros El Sachat\*,[1], Peng Xiao[1,2], Davide Donadio[3,4], Frédéric Bonell[5], Marianna Sledzinska[1], Alain Marty[5], Céline Vergnaud[5], Hervé Boukari[5], Matthieu Jamet[5], Guillermo Arregui[1], Zekun Chen[3,4], Francesc Alzina[1], Clivia M. Sotomayor Torres[1,6], Emigdio Chavez-Angel\*,[1]*

[1] Catalan Institute of Nanoscience and Nanotechnology (ICN2), CSIC and BIST, Campus UAB, Bellaterra, 08193 Barcelona, Spain

[2] Departamento de Física, Universidad Autónoma de Barcelona, Bellaterra, 08193 Barcelona, Spain

[3] Department of Chemistry, University of California, Davis, California 95616, United States

[4] Ikerbasque, Basque Foundation for Science, E-48011 Bilbao, Spain

[5] Université Grenoble Alpes, CNRS, CEA, Grenoble INP, IRIG-Spintec, 38054 Grenoble

[6] ICREA, Passeig Lluis Companys 23, 08010 Barcelona, Spain

\*Corresponding authors: alexandros.elsachat@icn2.cat, emigdio.chavez@icn2.cat

# 1. Reflection high energy electron diffraction, X-ray diffraction and atomic force microscopy measurements

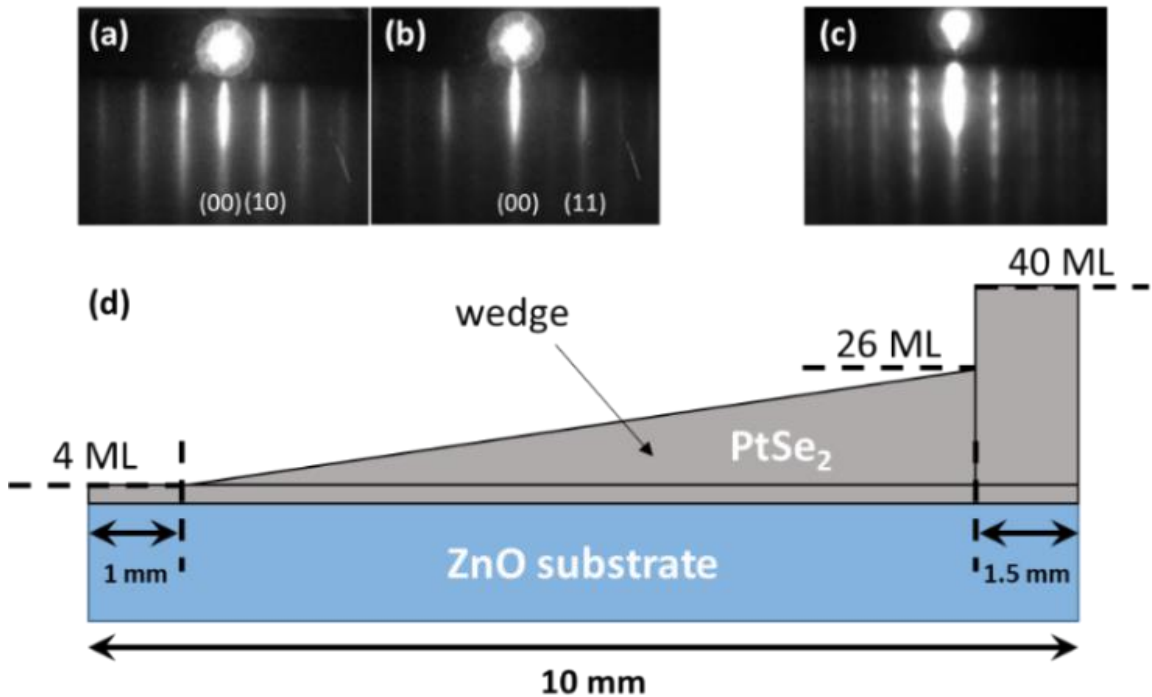

**Supplementary Figure S1. Reflection high energy electron diffraction (RHEED) measurements and the final structure of the PtSe₂ wedge sample.** **(a)** and **(b)** RHEED patterns of 4 ML PtSe$_2$ grown by co-evaporation along two different azimuths respectively. **(c)** RHEED pattern of the polycrystalline PtSe$_2$ film. Only one azimuth is shown since it is isotropic. **(d)** Schematic drawing in cross-section of the final structure.

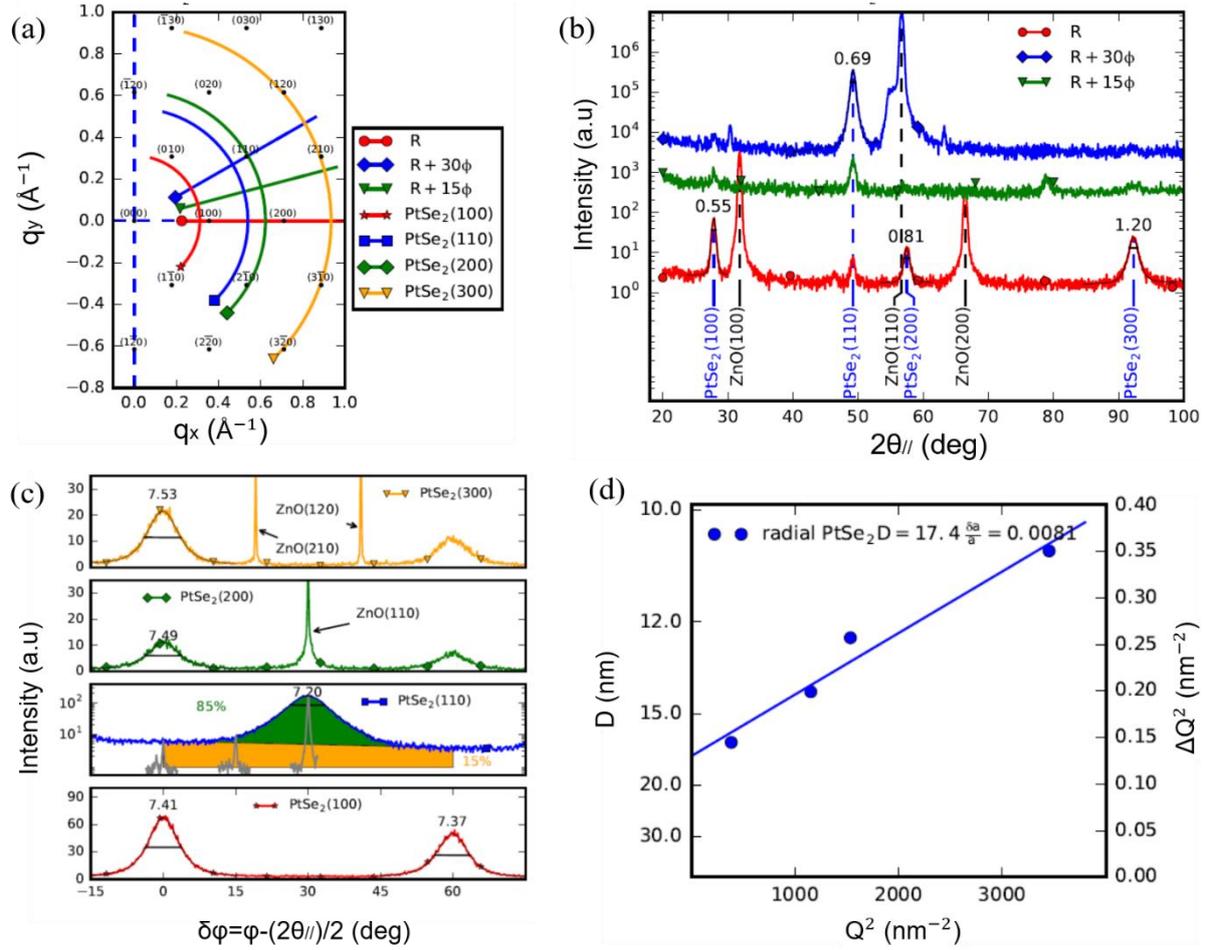

**Supplementary Figure S2. X-ray diffraction measurements in PtSe$_2$**. **(a)** Schematics of the different scans in the reciprocal space (radial and azimuthal) performed by grazing-incidence x-ray diffraction, **(b)** XRD radial scans showing the crystalline nature of the PtSe$_2$ film. **(c)** Azimuthal scans of the PtSe$_2$ Bragg peaks, from which we obtain the in-plane mosaic spread (see numbers on top of the peaks corresponding to the FWHM in degrees) **(d)** The momentum transfer Q for each radial Bragg diffraction peak can be calculated using $Q = 4\pi/\lambda \sin(\theta_{//})$, and the radial and azimuthal peak momentum dispersions using $\Delta Q_{rad}=\Delta(2\theta)/2 \; 4\pi/\lambda \cos(\theta_{//})$, where $\Delta(2\theta)$ are the corresponding FWHM. Plotting $\Delta Q^2$ as a function of $Q^2$ for the radial and azimuthal scans, one can determine the domain size D, the in-plane lattice parameter distribution $\Delta a/a$, from the intercept and the slope of the radial and azimuthal scans respectively using the linearized expressions:[1] $\Delta Q^2_{rad} = (2\pi/D)^2 + Q^2 (\Delta\alpha/\alpha)^2$.

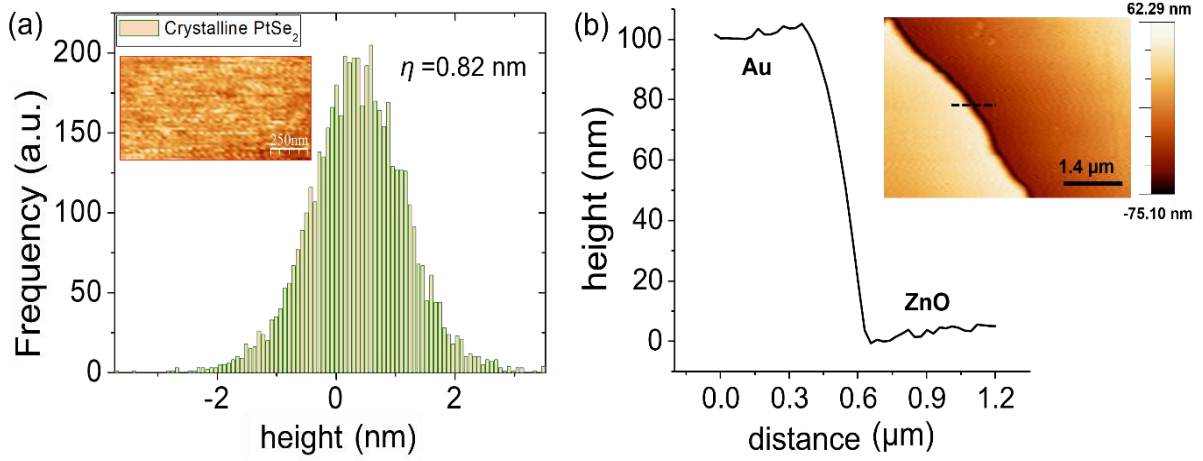

**Supplementary Figure S3. AFM measurements. (a)** Histogram of the height distribution (surface roughness) measured by AFM for crystalline PtSe$_2$. The root mean square roughness (RMS), $\eta$, is 0.82 nm. The inset shows an AFM image of the crystalline PtSe$_2$ wedge sample. **(b)** AFM topography profile of the deposited Au transducer on ZnO substrate taken from the black dashed line depicted in the AFM image (inset in (b)).

## 2. FDTR sensitivity analysis

We quantify the phase sensitivity to multiple thermal properties in a similar manner to that of Schmidt et al.[2] The calculated phase sensitivity ($-V_{in}/V_{out}$) to multiple parameters ($k_z$, $C_v$, anisotropy, $G_1$ and $G_2$) as a function of thickness and modulation frequency for the case of Au/PtSe$_2$/ZnO stacks is shown in Fig. S4(a-e). For the sensitivity analysis, we fit the experimental thickness dependence of the cross-plane thermal conductivity ($k_z$) using the Fuchs-Sondheimer model considering only cross-plane heat transport described by:[3]

$$\frac{k_{film,\,z}}{k_{bulk,\,z}} = S(\chi) = 1 - \chi(1 - \exp[-1/\chi]) \tag{S1}$$

where $k_{bulk, z}$ is the cross-plane thermal conductivity of the film (bulk) sample, S($\chi$) is the suppression function, $\chi$ is the Knudsen number $\chi = \Lambda_{bulk}/d$, $\Lambda_{bulk}$ is the bulk MFP (120 nm from the best fit) and $d$ is the thickness of the film. This fit allows us to have a continuous function to plot the sensitivity as a function of excitation frequency and sample thickness. The degree of anisotropy or the ratio between in-plane ($k_r$) and cross-plane ($k_z$) thermal conductivities was estimated from the bulk measurements $k_r/k_z = 9$. It was assumed to be constant for all the films.

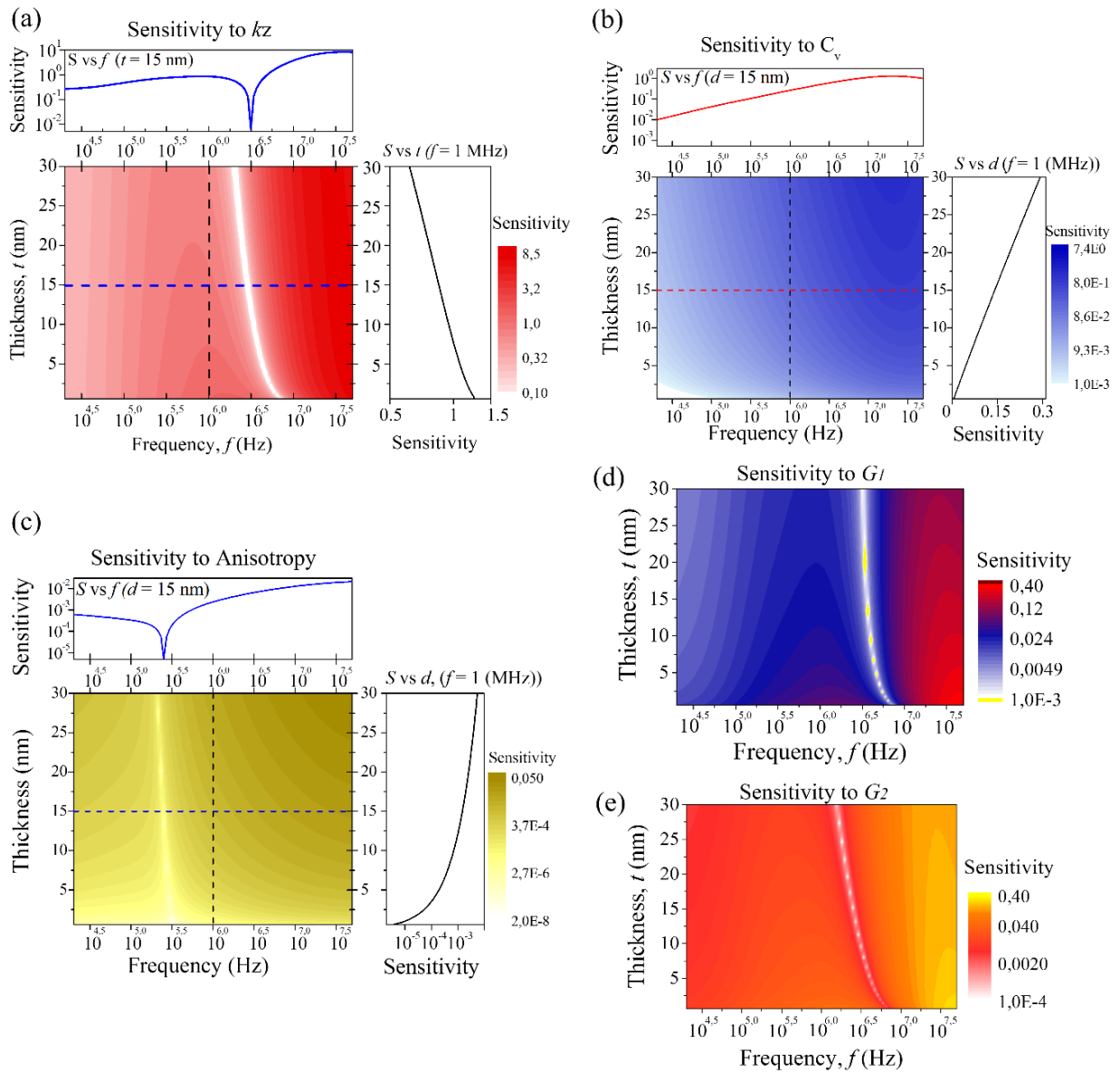

**Supplementary Figure S4. Phase sensitivity analysis for thin PtSe$_2$ films.** Calculated phase sensitivity to **(a)** $k_z$, **(b)** $C_v$, **(c)** anisotropy, **(d)** $G_1$ and (e) $G_2$ as a function of thickness and modulation frequency for the case of Au/PtSe$_2$/ZnO stacks.

The sensitivity analysis is shown over a frequency range of 20 kHz–40 MHz, where we observed that depending on the frequency range the phase sensitivity to a given parameter varies. For example, the sensitivity to in-plane transport for the thin films is almost zero and it is dominant for cross-plane transport (see Fig. S4c). Moreover, from Fig. S4d,e we observe that the sensitivity of the recorded phase signal to $G_1$ and $G_2$ is relatively low.

However, in the case of bulk $PtSe_2$ the sensitivity to cross-plane and in-plane transport is high almost in the entire frequency range (see Fig. S5), thus it is possible to extract the anisotropy ratio between in-plane ($k_r$) and cross-plane ($k_z$) thermal conductivity from a data set.

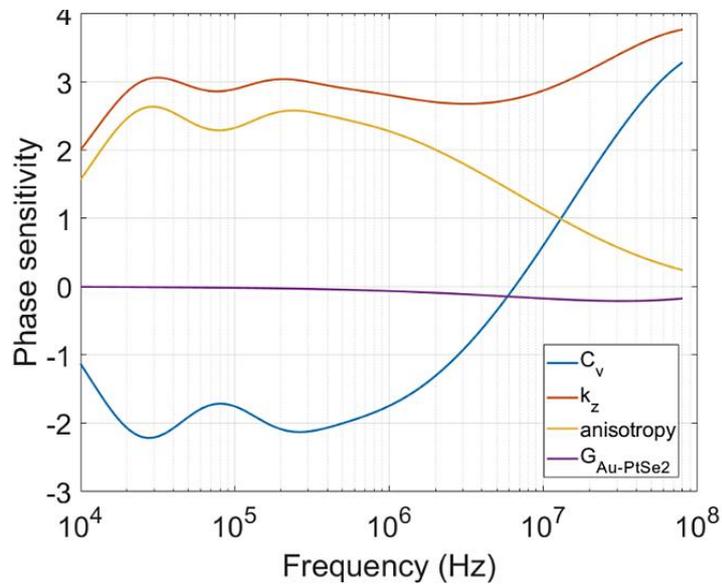

**Supplementary Figure S5. Phase sensitivity analysis for bulk $PtSe_2$.** Calculated phase sensitivity to $k_z$, $C_v$, anisotropy and $G_{Au-PtSe2}$ as a function of the modulation frequency for the case of bulk $PtSe_2$ crystal.

## 3. Spot size measurements

The spot size for each measurement was determined by using the knife's edge method. For the measurements we use the transducer layer as sharp edge and we measured the intensity of the reflected light as a function of the stage position. The beam intensity as a function of the translation distance was fitted to an error function curve[4] and the $1/e^2$ radius of this curve was taken as the laser spot radius.

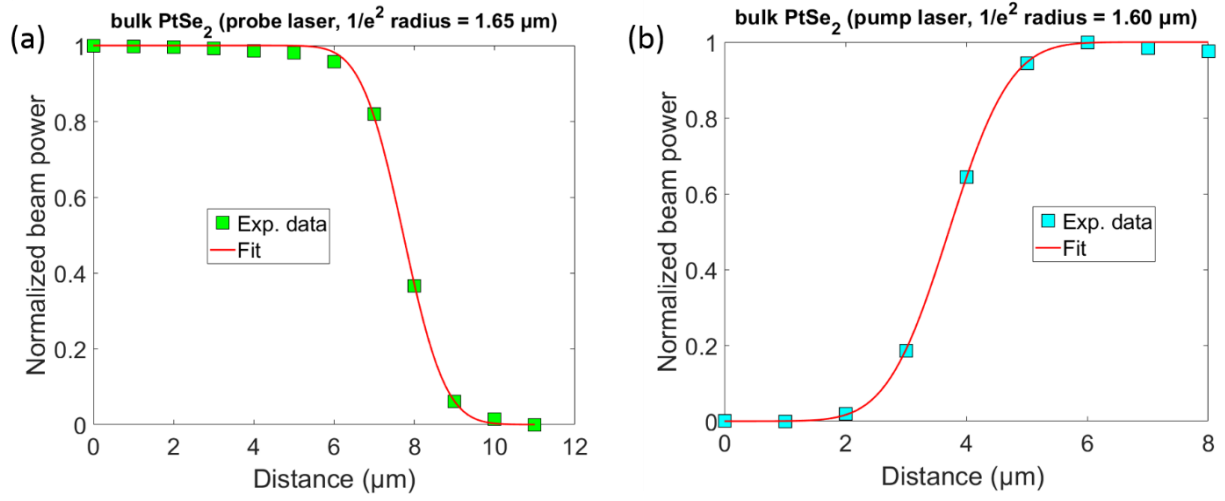

**Supplementary Figure S6. Spot size measurements.** Typical examples of measured **(a)** probe and **(b)** pump spot sizes in bulk PtSe$_2$ crystal using the knife's edge method.

## 4. Three omega measurements

The measurement of the thermal conductivity of the ZnO substrate was carried out by using the well-known three-omega (3ω) method.[5,6] The 3ω-heater strip consisted in a 100 nm-thick gold wire patterned on a 0.5 mm thick ZnO substrate by photolithography and electron beam physical vapour deposition (EBPVD). The width of the heating line was defined as $2b = 12$ μm and the length as $l = 1.3$ mm, the latter considered as the distance between the voltage (inner) pads. The width was estimated by averaging several microscopic images of the heater lines. The thermal conductivity of the substrate was obtained by solving the transient heat conduction equation for a finite width line heater, deposited onto semi-infinite surface of a film-on substrate system. The temperature rise is given by the following equation:

$$\Delta T = \frac{P}{lk\pi} \int_0^\infty \frac{\sin^2(xb)}{\sqrt{x^2 + iq^2}} dx \qquad (S2)$$

where P is the applied power (20 mW), $q = 1/\lambda = \sqrt{2\omega/\alpha}$ is the inverse of the thermal penetration depth ($\lambda$), with $\alpha$ the thermal diffusivity and $\omega$ the excitation frequency ($\omega = 2\pi f$), and $k$ is the thermal conductivity of the material. Finally, the thermal conductivity of the substrate was estimated by fitting the experimental temperature rise of the 3ω signal (grey, orange and green dots Fig. S7a) using Eq. (S2).

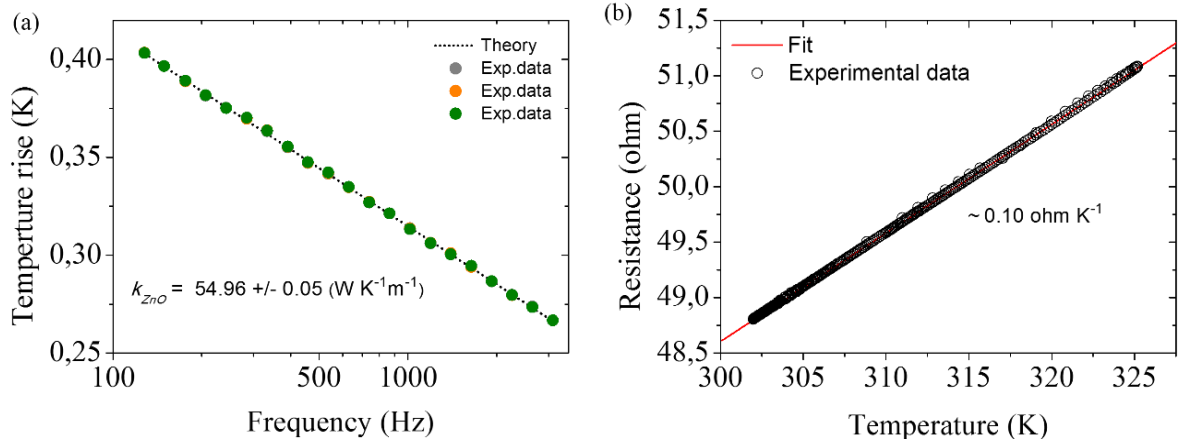

**Supplementary Figure S7. Three omega measurements. (a)** Temperature rise of the 3ω-heater vs excitation frequency for a 0.5 mm thick ZnO substrate. **(b)** Typical temperature dependence of the electrical resistivity of a 3ω-heater of the same system.

## 5. Raman thermometry measurements

To validate the thermal conductivities obtained by the FDTR measurements, Raman thermometry analysis were carried out in 2, 3 and 4 layer-thick PtSe$_2$ supported samples. Raman thermometry is a contactless and steady state technique for measuring thermal conductivity. It is based on probing the local temperature using the shift of Raman mode as a thermometer, previously calibrated in a cryostat at a pressure of $10^{-3}$ mbar. We used the excitation laser (532 nm) as both the heat source and the temperature sensor simultaneously. The local heating was controlled by varying the incident laser power and the thermal conductivity was extracted by solving the heat equation using finite element simulation method (COMSOL). In the simulations, the heat power and thermal properties of the sample were given as inputs, and the resulting temperature profile was calculated. Then, by adjusting the simulated temperature profile to the measured one, the cross-plane thermal conductivity was obtained.

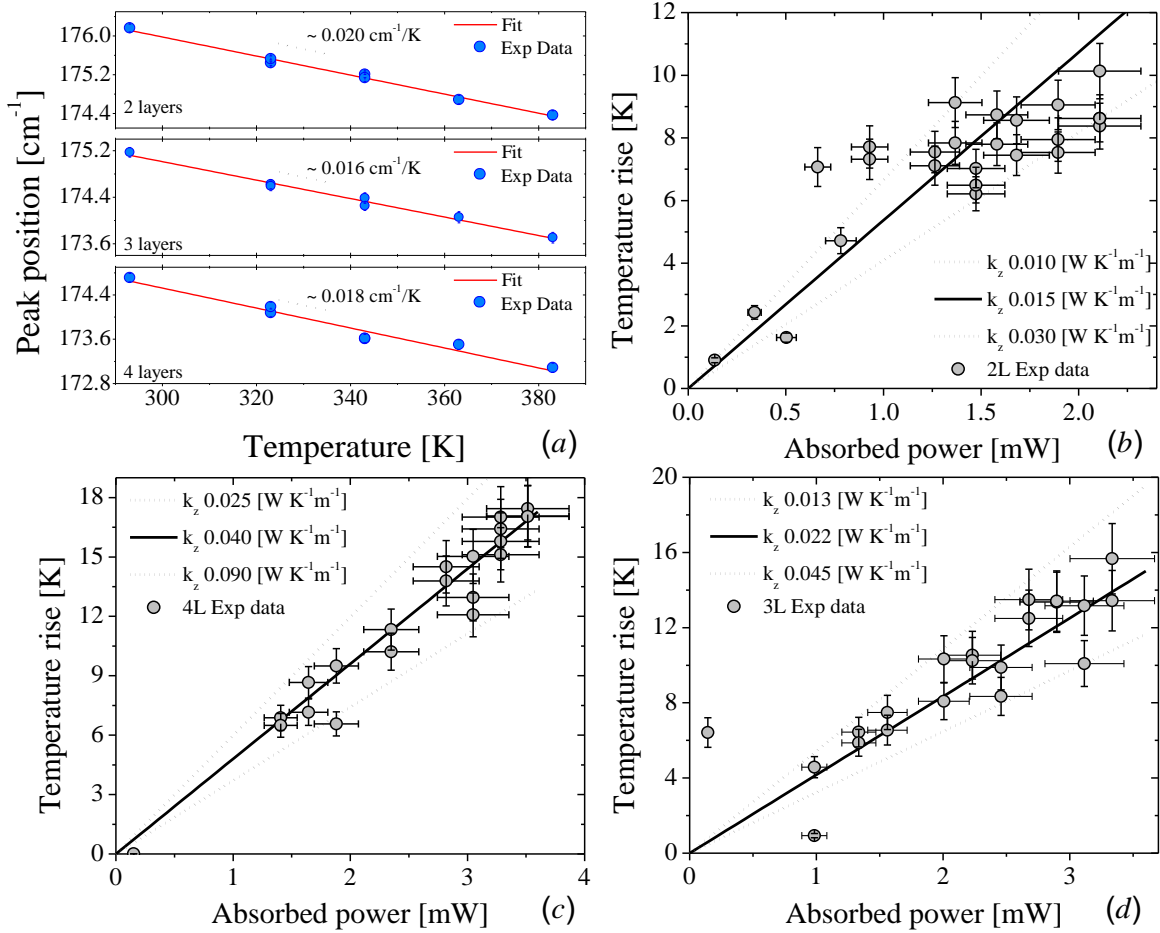

**Supplementary Figure S8. Raman thermometry measurements. (a)** Temperature dependence of the $E_g$ Raman active mode peak position. **(b)-(d)** Experimental temperature rise as a function of the absorbed power.

The absorbed power was estimated by using the experimental absorption reported by Chen et al.[7] and extrapolated to our experimental data considering a 10% deviation (x-axis error bar). For the FEM simulations, we used a Gaussian heat source given by:

$$Q = \frac{2AP_l}{\pi d\sigma^2} \exp\left[\frac{-2r^2}{\sigma^2}\right] \tag{S3}$$

where $A$ is the absorption coefficient, $P_l$ is the incident power, $d$ is the sample thickness and $\sigma$ is the $1/e^2$ radius measured in-situ using the knife's edge method. The temperature rise in the film ($T_m$) is averaged considering the real spatial temperature distribution given by:

$$T_m = \frac{\int_0^\infty rT(r)\exp\left[\frac{-2r^2}{\sigma^2}\right]dr}{\int_0^\infty r\exp\left[\frac{-2r^2}{\sigma^2}\right]dr} \tag{S4}$$

A comparison of the measured $k_z$ by Raman thermometry and FDTR is shown in table S1.

| Sample | Measured thermal conductivity (WK$^{-1}$m$^{-1}$) | |
|---|---|---|
| | Raman thermometry | FDTR |
| 2 layers PtSe$_2$ | 0.015 | 0.04 |
| 3 layers PtSe$_2$ | 0.040 | 0.06 |
| 4 layers PtSe$_2$ | 0.09 | 0.10 |

**Supplementary Table S1. Measured cross-plane thermal conductivity obtained by Raman thermometry and FDTR.**

## 6. Fourier transform from ASOPS data

Figure S9 shows the Raman spectra and fast Fourier transform of the layered breathing modes for different film thicknesses. From the FFT spectra we can observe a second peak located at higher energy than the LBM. It means that the experimental data should be fitted with two-exponential model as is also shown in the work of Chen et al.[7]. The origin of this second peak is associated to a Raman inactive standing wave produced by the interaction of the thin film with the substrate.

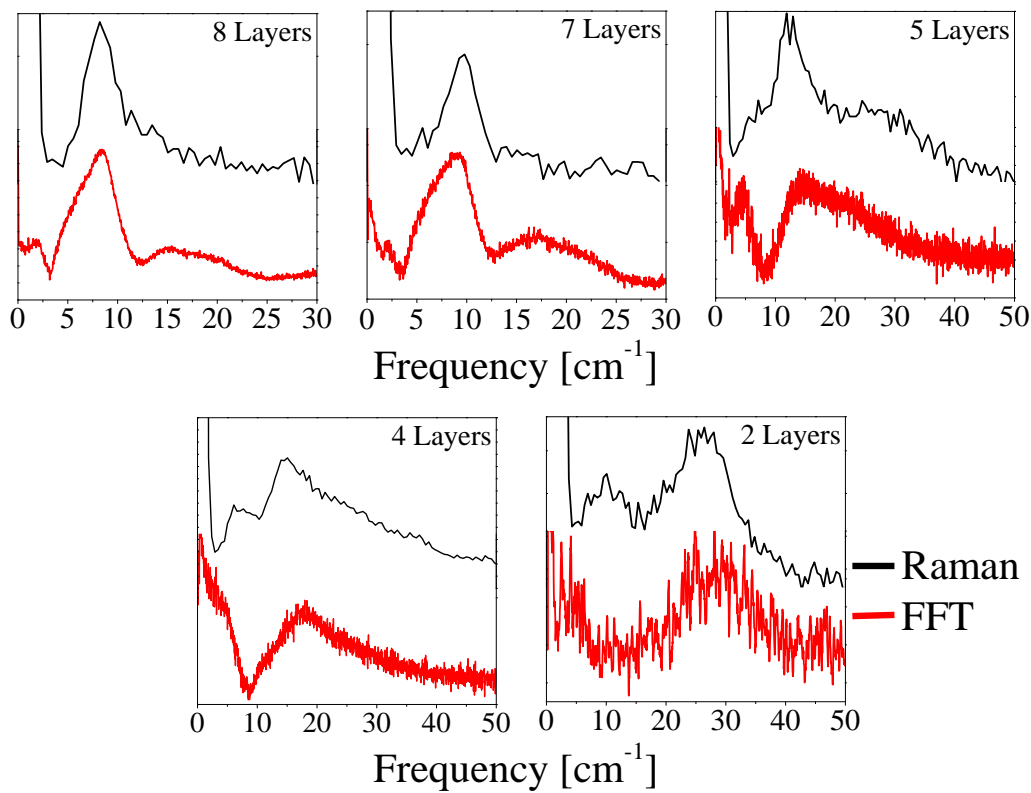

**Supplementary Figure S9. Fast Fourier transform of the layered breathing modes.** Layer-dependent low breathing modes measured by Raman spectroscopy (black line) and fast Fourier transform (FFT) of ASOPS signals (red lines).

## 7. DFT Calculation

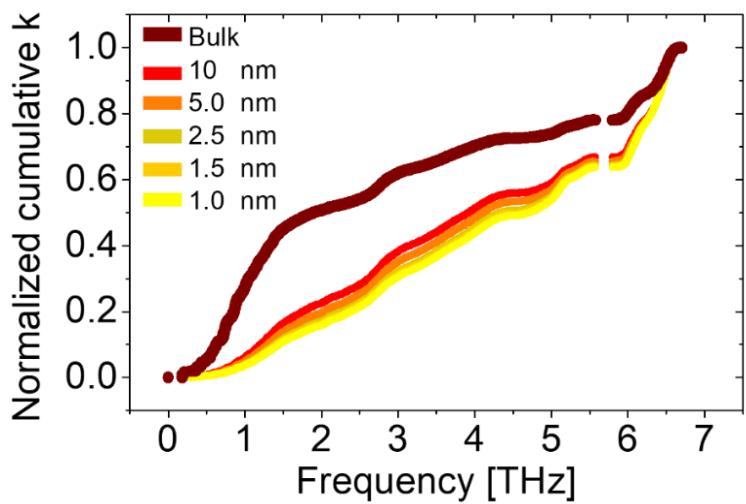

**Supplementary Figure S10. DFT calculations.** Normalized thermal conductivity accumulation as a function of the phonon frequency for thin films of thicknesses ranging from 1 to 10 nm and bulk PtSe$_2$.